\documentclass[aps,prl,twocolumn,amsmath,reprint,amssymb,showpacs]{revtex4-1}
\usepackage{graphicx}
\usepackage{dcolumn}
\usepackage{bm}
\usepackage{hyperref}
\usepackage{subfigure}
\usepackage{url}
\usepackage{ulem}
\usepackage{color}
\newcommand{\julien}[1]{\textcolor{black}{#1}}

\begin{document}

\title{Dynamic wrinkling and strengthening of an elastic filament in a viscous fluid}

\author{Julien Chopin}
\email{jchopin@clarku.edu}
\affiliation{Laboratoire Sciences et Ing\'enierie de la Mati\`ere Molle, PSL Research University, UPMC
Univ Paris 06, ESPCI Paris, CNRS, 10 Rue Vauquelin, 75231 Paris Cedex 05, France}
\author{Moumita Dasgupta}
\altaffiliation{Present Address: Department of Physics  \& Astronomy, Amherst College, Amherst, MA 01002}
\author{Arshad Kudrolli}
\email{akudrolli@clarku.edu}
\affiliation{Department of Physics, Clark University, Worcester, Massachusetts 01610, USA}
\date{\today}

\begin{abstract}
\julien{We investigate the wrinkling dynamics of an elastic filament immersed in a viscous fluid submitted to compression at a finite rate with experiments and by combining geometric nonlinearities, elasticity, and slender body theory. The drag induces a dynamic lateral reinforcement of the filament leading to growth of wrinkles that coarsen over time. We discover a new dynamical regime characterized by a timescale with a non-trivial dependence on the loading rate, where the growth of the instability is super-exponential and the wavenumber is an increasing function of the loading rate.  We find that this timescale can be interpreted as the characteristic time over which the filament transitions from the extensible to the inextensible regime. In contrast with our analysis with moving boundary conditions, 
Biot's analysis in the limit of infinitely fast loading leads to rate independent exponential growth and wavelength.}
\end{abstract}

\maketitle
Slender structures embedded in complex fluids which buckle and fold as a result of mechanical compression are commonly found as in F-actin and microtubules in cell mechanics~\cite{Gardel2004,Chaudhuri2007,Jiang2008} flagella in swimming organisms~\cite{Powers2010,Goldstein2006,Son2013}, fibers in paper processing~\cite{Lindner2012}, and the earth's crust in orogenesis~\cite{Biot1961}. A classical result dating back to Euler states that a thin sheet or filament will buckle under axial loading above a critical strain which is proportional to the square of the mode number and the square of the ratio of its thickness to length~\cite{timoshenko1940strength}. While buckling typically occurs in the fundamental mode corresponding to the lowest strain, higher modes can occur depending on the constraints along the filament which may be static or dynamic in nature~\cite{Chopin2013,Chopin2015,miller2015buckling,lagrange2016wrinkling,Audoly2005,Vermorel2007,gladden2005dynamic}. 
Although theoretical analysis of the problem are numerous, there are few experimental systems allowing close comparison with predictions. \julien{Surprisingly, since the work by Biot, \textit{et al.}~\cite{biot1961experimental}, where the phenomena was first demonstrated, no careful experimental studies have systematically examined the wrinkling dynamics of an isolated elastic filament in a viscous Newtonian fluid.} 

Traditional analysis to the wrinkling observed in elas-
tic  filaments  consider  linear  stability  analysis  with  instantaneous loading which can be an oversimplification in
many situations ~\cite{Biot1961,Huang2001,Jiang2008,Li2008,kodio2016lubricated}. For example, \julien{the folding of sedimentary layers,} the buckling of membranes, the motion of living systems or actuated membrane involves the change of distance between material points in response to a stimuli (stress, light, pH, temperature) whose dynamics is set externally\julien{~\cite{budd2000approximate,forterre2005venus,stuart2010emerging,jager2000microfabricating,osada1992polymer,chen1995graft,kim2012designing,camacho2004fast,van2009printed,yu2003photomechanics}}. Additionally, in-plane and out-of-plane modes experience dynamics with a timescale highly dependent on the surrounding environment as in air~\cite{gladden2005dynamic}, intra-cellular medium~\cite{Brangwynne2007}, or a viscous fluid~\cite{wiggins1998trapping}.  In this context, the interplay between the dynamics of an external stimuli and the dynamics of the elastic modes have been largely overlooked. Here, we examine a model system that reveals the emergence of a new timescale for the growth of wrinkles that explicitly depends on the loading rate. The precise understanding of this interplay is important for the development of fast reacting metamaterials and the tuning of the rheological response of polymers.

Experiments were performed with an elastic filament clamped at both ends and immersed in a container filled with a viscous fluid. The filament has a length $L = 92\,$mm, width $W = 3.0\,$mm and thickness $h = 0.30\,$mm and is composed of vinylpolysiloxane with a Young's modulus $E= 1\,$MPa, and Poisson's ratio $\nu \approx 0.5$. The bending and stretching moduli are given by $B = Eh^3W/12$ and $K = EhW$ respectively. 
An anisotropic filament cross section was chosen to have a well defined plane in which the filament buckles, simplifying data acquisition and analysis. One clamped end of the filament is attached to a motorized translating stage which can be moved through a displacement $u$ which varies between $u_0 < 0$ when the filament is under tension, and $u_f > 0$ when the filament is under compression  with speed $V_e$ in the range $5\times10^{-3}$mm/s -- $10\,$mm/s. The fluid composed of a glycerol-water mixture is prepared so that the filament is neutrally buoyant and has a dynamic viscosity $\eta= 0.9\,$Pa\,s at 25$^{\circ}$C. The viscous drag coefficient is given by $\mu \approx 4\pi \nu /\log(L/W)$ \cite{batchelor1970slender,Powers2010}. Unless otherwise stated, distances and time are normalized by $L$ and a time scale $\tau = \mu L^2/K=3\times10^{-2}$\,s given by a balance between viscous drag and elastic forces. With this rescaling, the non-dimensional speed $V = \tau V_e /L$ ranges between $2\times 10^{-6}$ and $3\times 10^{-3}$.

The filament is imaged with a $1200 \times 1200$ pixel camera at 1500\,fps by placing a diffused light source on the opposite side of the container which 
allows a good contrast for subsequent image analysis. We then obtain the filament deflection $w$ to within $0.15 h$ as a function of the non-dimensionalized coordinate along the longitudinal direction  $x$ and time $t$. We calculate the bending content $\kappa(t) = h^2 \langle (w_{,xx}(x,t))^2 \rangle $, where $\langle ..\rangle$ denotes average over the length of the filament, and the averaged strain normalized by $L$ given by
\begin{eqnarray}
\gamma(t) =  -u(t) + 1/2 \int_0^1 w_{,x}^2(x,t) dx\,,
\label{eq:gamma}
\end{eqnarray}
This expression is obtained by integrating the 1D nonlinear strain $\gamma(x,t) = u_{,x}(x,t) + 1/2 (w_{,x}^2(x,t))$, where $u(x,t)$ is the local in-plane axial displacement~\cite{landau1986course}. The evolution of the stretching and bending energies stored in the filament can be then quantified by $\gamma(t)$ and $\kappa(t)$, respectively.

\begin{figure}
    \centering
    \includegraphics[width = 8.5cm]{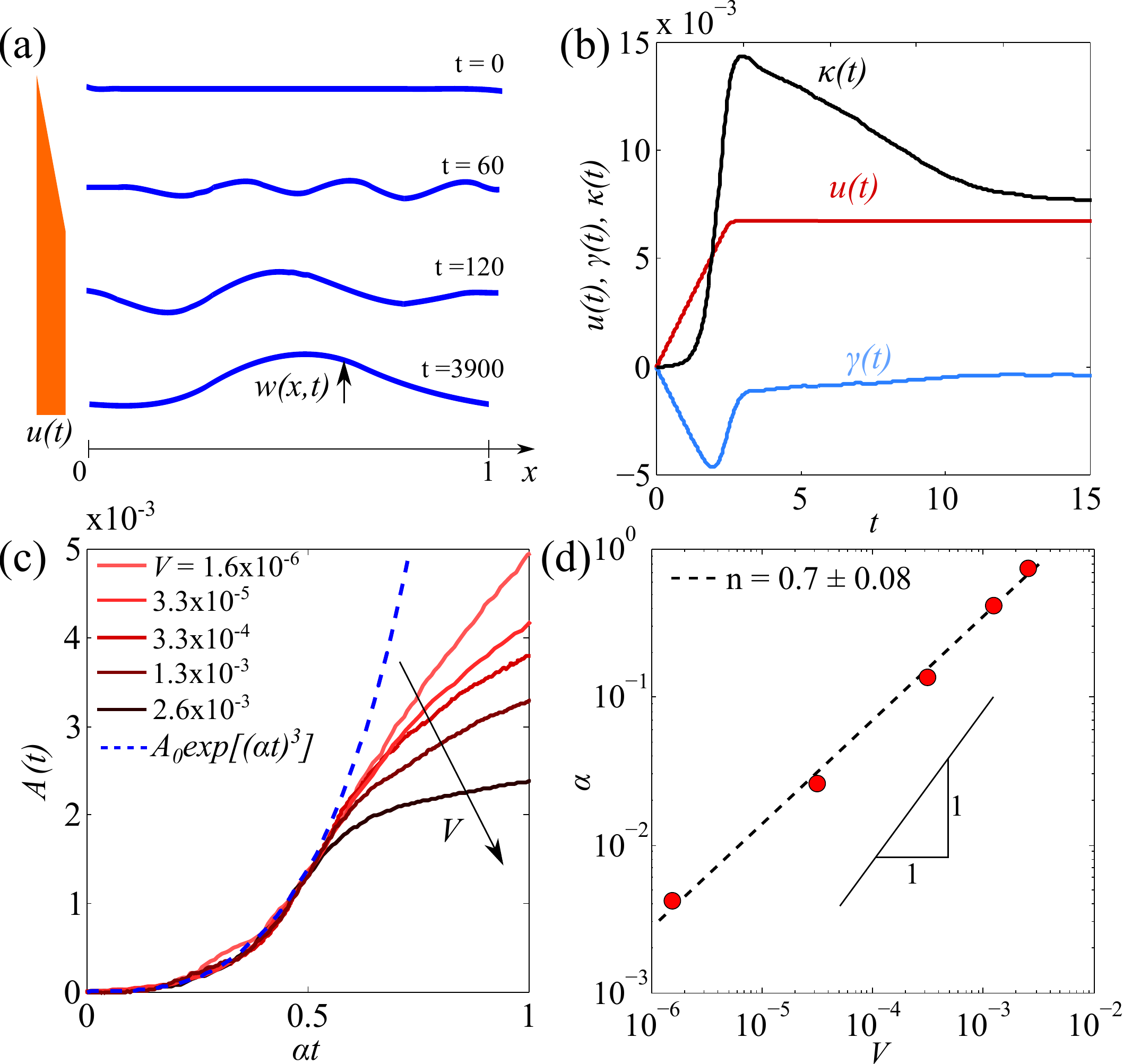}
    \caption{(a) Snapshots of the measured filament deflection $w(x,t)$ at various times $t$ with the applied strain as a function of time illustrated by the bar on the left ($u_f = 7\times10^{-3}$). Wrinkles emerge during loading followed by coarsening till the fundamental buckling mode is reached. (b) The measured strain $\gamma$ is observed to relax rapidly while the ribbon is being compressed. In contrast, the bending content $\kappa(t)$ associated with the coarsening dynamics relaxes much slower while the ends of the filament are held with a prescribed strain. (c) The initial evolution of $A(t)$ for various $V$ is observed to collapse on to a single curve using a fitting parameter $\alpha$. (d) $\alpha$ is observed to scale sub-linearly as a function of $V$ consistent with Eq.~\ref{eq:alpha}.}
    \label{Fig1}
\end{figure}

Fig.~\ref{Fig1}(a) shows the measured profile at various times while the filament is loaded as denoted by the bar on the left. The filament is observed to undergo a wrinkling instability during the loading phase ($t<t_f = u_f/V$) with a wavelength which is approximately a third of the length of the ribbon. \julien{Wavevector larger that the fundamental buckling mode indicates a higher critical buckling load, hence a dynamical strengthening induced by the viscous medium.} Shortly thereafter, the ribbon is observed to undergo spontaneous coarsening till the wavelength reaches the filament length, and the shape corresponds to the fundamental buckling mode calculated using time-independent Euler analysis~\footnote{see Supplementary Documentation for a movie of the shape evolution, the calculation of the Euler buckling modes, and the comparison of the fundamental mode with the final shape shown of the filament.}.  
 We obtain the evolution of $\gamma$ and the bending content $\kappa$, 
and plot $u(t)$, $\gamma(t)$, and $\kappa(t)$ in Fig.~\ref{Fig1}(b) focusing on a time interval over which most of the initial buckling and coarsening occurs. At the onset of instability, $\gamma(t)$ and $\kappa(t)$  are observed to increase rapidly showing the increase of the bending energy and the relaxation of the compression. From the plot, we observe that the rate of change of $\gamma(t)$ and $\kappa(t)$ starts to decrease well before compression is stopped. Although $\gamma(t)$ is found to vanish much faster than $\kappa(t)$, stretching is expected to play an important role at onset. We can then anticipate that a wrinkling growth model is likely to include small but finite extensibility and bending rigidity.

We obtain the root mean square amplitude $A(t) = \sqrt{\langle w^2(x,t) \rangle}$ from the measured $w(x,t)$ to understand the growth of the buckling modes. Fig.~\ref{Fig1}(c) shows a plot of $A(t)$ as a function of time scaled with a parameter $\alpha$. This parameter is chosen so that the initial growth of $A(t)$ collapses onto a single curve before a time when its rate of increase starts to decrease. Plotting $\alpha$ as a function of $V$ in log-log scale in Fig.~\ref{Fig1}(d), we find that the data can be described by the function  $\alpha \sim V^n$ with $n = 0.7 \pm 0.08$. This form is observed to be significantly different from a linear scaling if the dependence on time was captured simply by $V$.

To develop a model of this non-trivial scaling, we next consider the spatio-temporal variation of the normal deflection $w(x,t)$ and the axial strain averaged over the length given by $\gamma(t)$. We assume that axial force balance develops rapidly in our system because of the homogeneous development of wrinkles along the length of the filament in contrast with observations with very viscous fluids which show wrinkling localized near the moving ends~\cite{Gosselin2014}. Under these conditions, the axial tension in the filament $T$ just depends on time and is given by $T(t) = K \gamma(t)$. 
\julien {Reynolds number is typically low, $Re = \rho \dot{A} W/\eta \sim 10^{-1}$ where $\rho \sim 10^{3}$ kg/m$^{3}$ is the fluid density and $\dot{A} \sim 10^{-1} $ m/s (see Fig.~\ref{Fig1}(c)). }Then, neglecting inertia and balancing drag, bending and stretching forces, the non-dimensional equation of evolution for $w(x,t)$ is given by 
\begin{eqnarray}
\dot{w} &=& -\frac{B}{K}w_{,xxxx} + \gamma(t) w_{,xx}\,.
\label{eq:deflec}
\end{eqnarray}
The corresponding boundary conditions are $w(0,t) = w(1,t)= 0$, $w_{,x}(0,t)=  w_{,x}(1,t) =0$, $u(1,t) = 0$, and
\begin{eqnarray}
u(t) &\equiv& u(0,t) = 
\begin{cases}
    Vt,& \text{if } t<{u_f}/{V}.\\
    u_f,              & \text{otherwise}.
\end{cases}\, 
\label{eq:BC}
\end{eqnarray}
\begin{figure}[t]
    \centering
    \includegraphics[width = 8.5cm]{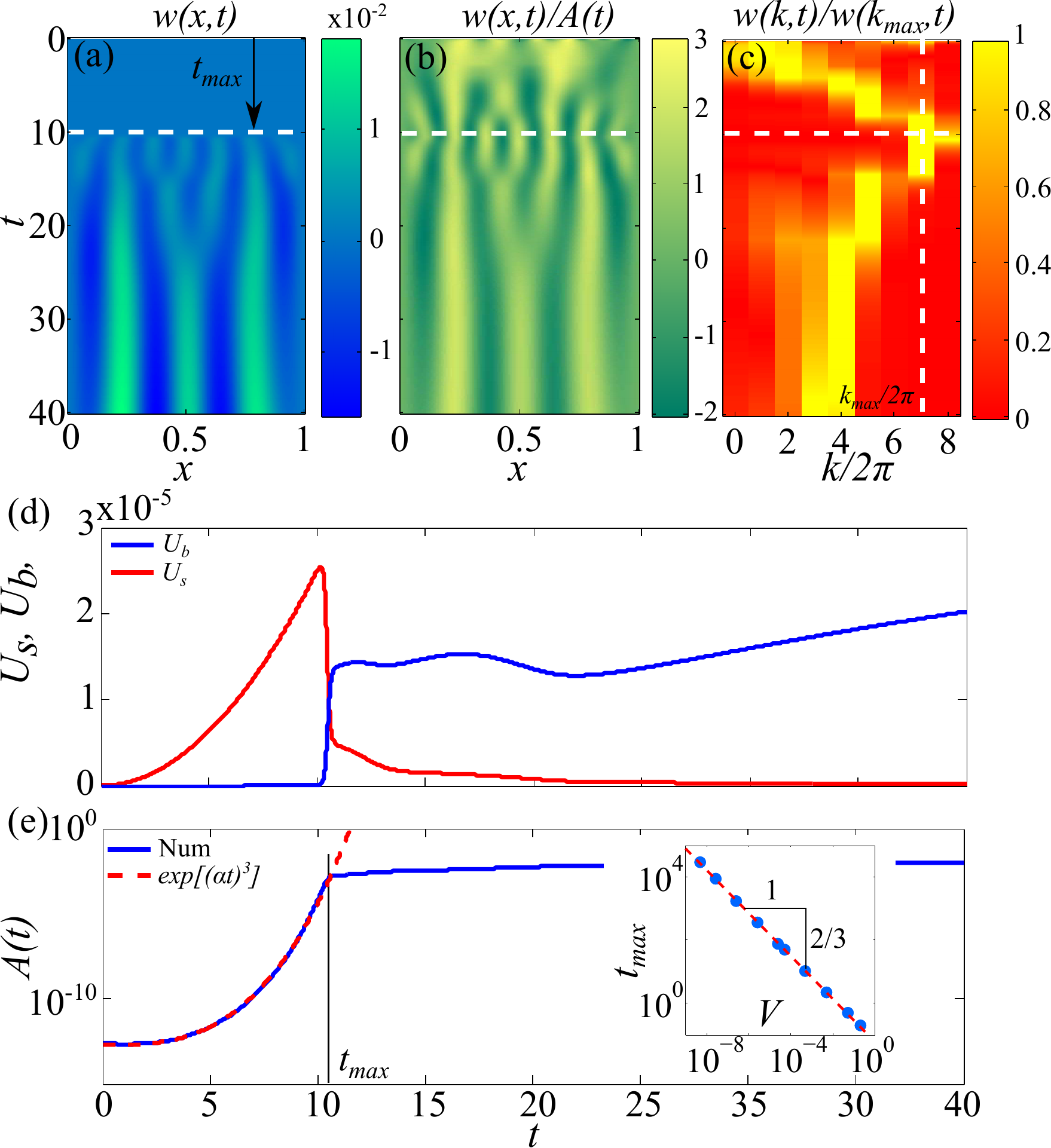}
    \caption{(a) A map of $w(x,t)$ for $V = 5.0\times 10^{-4}$ and $u_f = 0.1$. Compression stops at time $t_f = 200 \gg t_{max} = 10.5$. (b) Normalizing the same data as in (a) by $A(t)$ reveals the wrinkling dynamics before $t_{max}$. We observe an increase of the wavenumber until $t = t_{max}$ followed by a decrease. (c)  Map of the Fourier mode $w(k,t)$ showing the $k$ maximum reached at $t_{max}$.  Horizontal dashed line corresponds to $t_{max}$ in (a-c). (d) Temporal evolution of the stretching energy $U_s$ and the bending energy $U_b$ shows that $t_{max}$ is the crossover time between a stretching dominated regime (or extensible regime) and a bending dominated regime (or inextensible regime). (e) Semi-log plot of $A(t)$ shows a nonlinear growth where the amplitude can be fitted by $A_{fit}(t) = \exp{(\alpha t)^3}$. Inset: $t_{max}$ is observed to decay consistent with Eq.~\ref{tmax}.}
    \label{Fig2}
\end{figure}
\julien{The non-linear, integro-differential equation with moving boundary conditions given by Eqs.~\ref{eq:gamma}, \ref{eq:deflec}, and \ref{eq:BC} is solved numerically using the method of lines by discretizing the system and approximating partial differentials with finite differences in space~\cite{wouwer2014simulation}.} This results in a series of ordinary differential equations (ODEs) for the evolution of the deflection. Further, differentiating the strain given in Eq.~\ref{eq:gamma}, an additional equation for the evolution of the strain is obtained~\cite{kodio2016lubricated}. Then, by setting the initial conditions for the deflection with white noise, the equations are integrated forward in time using the ODE solvers in MATLAB~\cite{Note1}.

A sample initial spatio-temporal growth and coarsening of the wrinkles observed in the simulations by plotting $w(x,t)$ in Fig.~\ref{Fig2}(a) and $w(x,t)$ normalized by $A(t)$ in Fig.~\ref{Fig2}(b)~\cite{Note1}. 
 One can observe that the amplitude of the wrinkles increases rapidly and the wavelength appears to decrease till $t \approx 10$ before coarsening starts to develop and amplitude increases more slowly. To quantify the variation in wavelength, the temporal evolution of the Fourier modes as a function of $k$ and $t$ are plotted in Fig.~\ref{Fig2}(c). One observes that the peak in $k$ increases initially before decreasing confirming the trends noted in the evolution of $w(x,t)/A(t)$ in Fig.~\ref{Fig2}(b). Comparing the three plots, one also observes that the rapid increase in amplitude occurs at a time similar to time $t_{max}$ at which the wavenumber reaches a maximum $k_{max}/2\pi = 7$. 

We then measure the normalized stretching energy \julien{$U_s =\frac{1}{2} \int \gamma^2(x,t) dx$} and bending energy $U_b =  \frac{1}{2} \frac{B}{K} \int (w_{,xx}(x,t))^2 dx$. \julien{ As shown in Fig.~\ref{Fig2}(d), $t_{max}$ is found to be the crossover time from a stretching dominated extensible regime near onset of instability to a bending dominated inextensible regime during coarsening.}

\begin{figure}
    \centering
    \includegraphics[width = 6.5cm]{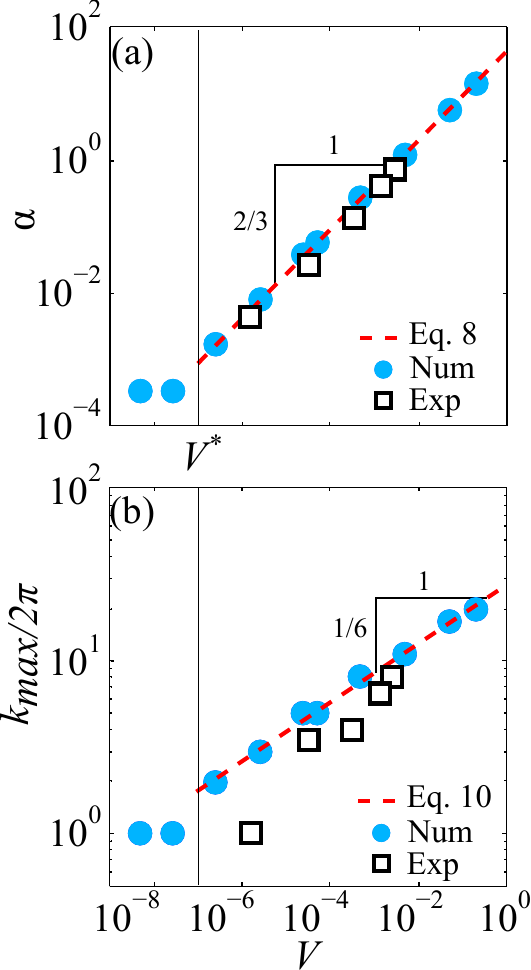}
    \caption{The observed evolution of (a) $\alpha$ and (b) $k_{max}$ with $V$ in experiments and numerical simulation are in excellent agreement with 
the derived scalings in the rate-dependent loading regime above $V^*$ indicated by a vertical line. }
    \label{Fig3}
\end{figure}

Fig.~\ref{Fig2}(e) shows a plot of $A(t)$ in semi-log scale which is observed to first increase faster than exponential followed by a much lower growth rate for $t>t_{max}$. We find that the initial growth is accurately fitted by $A_{fit}(t) = A_0 \exp{(\alpha t)^3}$ as shown in Fig.~\ref{Fig2}(e). Varying the compression speed $V$ in the range $10^{-8}$ and $10^{0}$, we measure the dependence of $t_{max}$, $k_{max}$, and $\alpha$ and plot the result in Fig.~\ref{Fig2}(e)(inset), Fig.~\ref{Fig3}(a), and Fig.~\ref{Fig3}(b), respectively. Because the observed trends are captured by the dashed lines, we find that $\alpha \sim 1/t_{max} \sim V^n$ with $n = 0.68 \pm 0.05$ in very good agreement with the experiment. Further, the wavenumbers measured in the experiments and in the simulations are also in agreement with $k_{max} \sim V^{0.17}$. Thus, we conclude that the simplified Eq.~\ref{eq:deflec} captures the overall evolution of the wrinkling patterns observed in the experiments. \julien{This allows us to next develop an understanding of the observed phenomena by performing a linear stability analysis around the planar undeflected configuration.}

Now, Eq.~\ref{eq:deflec} in Fourier space is given at linear order by
\begin{eqnarray}
\dot{\hat{w}}(k,t)=\left[u(t)k^2 -\frac{B}{K} k^4 \right] \hat{w}(k,t)\,,
\label{eqFourier}
\end{eqnarray}
where $\hat{w}(k,t)$ is the Fourier transform of $w(x,t)$. 
The general solution of this equation is simply $\hat{w}(k,t) = w_0 \, e^{\Gamma(k,t)}$ where $w_0$ is proportional to the amplitude of the white noise used as initial condition in the numerics and $\Gamma(k,t) = k^2\phi(t)-B/K k^4t $ with $\phi(t) = \int_0^t u(t')dt'$. The standard deviation of $w(x,t)$ is then
\begin{eqnarray}
A^2(t) &=& \frac{|w_0|^2}{2 \pi} \int_{-\infty}^{+\infty}e ^{2\Gamma(k,t)}dk\,, 
\end{eqnarray}
where we used the Parseval-Plancherel theorem. The integral can be evaluated using the Laplace method~\cite{budd2000approximate}. When $\phi(t) > 0$,  $\Gamma(k,t)$ reaches a maximum  for $k^2  =k^{*2}(t) = K/(2B) \phi(t)/t$. Therefore, 
\begin{eqnarray}
A(t) &\sim&\frac{|w_0|/2}{\pi^{1/4}} \frac{e ^{\Gamma^*(t)}}{ \phi^{1/4}(t)} \,,
\label{At}
\end{eqnarray}
where $\Gamma^*(t) = K/(4B) \phi^2(t)/t$. \julien{Eq.~\ref{At} is accurate for $t\gg1$}. Using the boundary conditions, we have  $\phi(t<u_f/V) = Vt^2/2$ and $\phi(t>u_f/V) = -u_f^2/(2V)+tu_f$. Thus, 
\begin{eqnarray}
\Gamma^* (t) &=& \frac{Ku_f^2}{4B}
\begin{cases}
    \frac{V^2t^3}{4u_f^2}\,, & \text{if }  t<\frac{u_f}{V}\,.\\
    \frac{u_f^2}{4V^2 t} -\frac{u_f}{V} + t\,, & \text{if } t>\frac{u_f}{V}\,.
\end{cases}\, 
\label{BigGamma}
\end{eqnarray}
Thus, we find that $A(t) \sim \exp\left[(\alpha t)^3 \right]$ during the loading phase. This super-exponential growth is consistent with the fits used to describe the experimental and numerical data, and where
\begin{eqnarray}
\alpha = \left( \frac{K}{8B}\right)^{1/3}V^{2/3}\,.
\label{eq:alpha}
\end{eqnarray}
This calculated scaling with $V$ is also in very good agreement with both the observed $n= 0.7 \pm 0.08$ in experiment in Fig.~\ref{Fig1}(d), and $n = 0.68$ in simulations in Fig.~\ref{Fig2}(e). \julien{Our analysis is accurate for the intermediate regime where $1 \ll t \ll 1/\alpha$ consistent with the experiment}. Further, the fastest mode has a wavenumber
\begin{eqnarray}
k^{*2}(t) &\approx& \frac{Vt}{4B/K}\,.
\end{eqnarray}
We cannot capture a time dependence of the wave vector near the initial growth because of the noise in the image detection, unlike $A(t)$ which is much cleaner. However, we can measure $k$ near coarsening, i.e. when non-linearities start to dominates. These non-linearities become dominant for large amplitude for $t>t_{max}$ where 
\begin{eqnarray}
t_{max} \sim 1/\alpha \sim V^{-2/3}.
\label{tmax}
\end{eqnarray}
Then, at time $t = t_{max}$ we have 
\begin{eqnarray}
k_{max}^2 = \left(\frac{K^2V}{8B^2} \right)^{1/3}.
\label{EqKmax}
\end{eqnarray}
The derived scaling of $k_{max} \propto V^{1/6}$ is consistent with the weak $V^{0.17}$ dependence seen in the numerics and the experiment shown in Fig.~\ref{Fig3}(b) for large enough $V$. \julien{Importantly, Eq.~\ref{EqKmax} shows that the dynamical strengthening is rate dependent, increasing with V, in contrast with previous studies.} At the lowest $V$ investigated, one observes that $k_{max}$ is constant and corresponds to the fundamental mode as can be expected. The crossover speed $V^*$ between wrinkling and buckling dynamics is derived using Eq.~\ref{EqKmax} and $k_{max} = 2\pi$ leading to $V^* \approx 8 (2\pi)^6 (B/K)^2 \sim 10^{-7}$ consistent with numerical and experimental results. Thus, our model explains quantitatively the dependence of the growth rate and the wavevector with the loading speed. 
\begin{figure}
    \centering
    \includegraphics[width = 6.5cm]{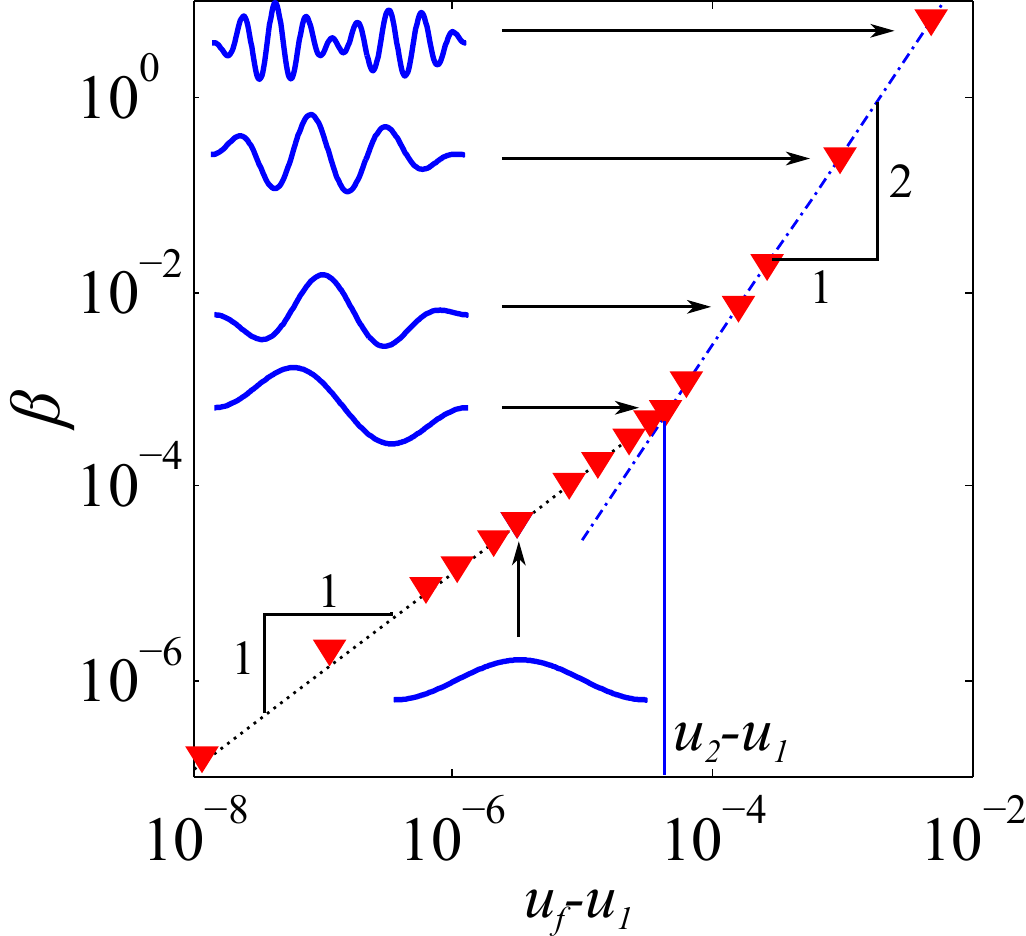}
    \caption{Evolution of $\beta$ with $u_f-u_1$ for an instantaneous compression. A crossover between a linear and quadratic trend occurs approximately at the critical compression for the first asymmetrical mode at $u_2 =  7.5 \times 10^{-5}$ yielding $u_2 -u_1 =  4.3 \times 10^{-5}$.}
    \label{Fig4}
\end{figure}

\julien{To contrast with the widely used sudden compression approximation, we consider the limit of large loading speed $V$ or, equivalently, long time ($t>u_f/V$). Using Eq.~\ref{BigGamma}, we recover the well-known exponential growth, $A(t) \sim e^{\beta t}$ where,
\begin{equation}
\beta = \frac{K u_f^2}{4B}\,,\,\,\, {\rm and} \,\,\,\, k^{*2} = \frac{Ku_f}{2B}\,,
\end{equation}
consistent with previous linear stability analysis~\cite{Biot1961}.} 
While the sudden compression regime is not reached in our experiment before coarsening develops, we can examine this regime by numerically solving Eq.~\ref{eq:deflec} by imposing an instantaneous compression $u_f$. Numerics confirms an exponential growth of the wrinkling amplitude with a growth rate $\beta$ which is shown in Fig.~\ref{Fig4} as a function of applied strain $u_f$. Interestingly, we find that the regime $\beta \sim u_f^2$ discussed \julien{in~\cite{Biot1961}} is only valid when $u_f  \gg u_1$, the strain corresponding to the buckling in the fundamental mode. Near threshold ($u_f \approx u_1$), we find another regime  characterized by $\beta$ linearly increasing with $u_f$. In this regime, the discreteness of the modes imposed by the boundary conditions plays a dominant role. When $u_1<u_f<u_2$, the only possible wave vector is $k = 2\pi$, yielding $\dot{w}(k,t)=4 \pi^2 (u_f - u_1) w(k,t)$, where $u_1 = 4 \pi^2 B/K$. Upon integration, we \julien{still} obtain an exponential growth with a growth rate linearly depending on $u_f-u_1$ as shown in Fig.~\ref{Fig4}. \julien{Therefore, the sudden compression approximation always leads to exponential growth and rate independent wrinkling dynamics inconsistent with our experiments.}

In conclusion, our study reveals that filaments subjected to a ramped loading show a super-exponential wrinkling growth in a regime where the filament is extensible which is qualitatively different than the growth derived when assuming a sudden loading. In particular, we find the wrinkles to grow as $\exp((\alpha t)^3)$ with a timescale $\alpha^{-1} \sim \beta^{-1/3} \dot{\gamma}^{2/3}$ which is a combination of the spontaneous timescale $\beta^{-1}$ of sudden compression and a timescale given by the loading rate $\dot{\gamma}$. 
\julien{This super-exponential growth is found to develop for time $t<t_{max}\sim \dot{\gamma}^{-2/3}$ before entering in an inextensible regime when coarsening occurs.} Our analysis based on our model experimental system thus advances a new time scale which is important to the mechanics of filaments which are subjected to time-dependent confining stresses.

\begin{acknowledgments}
We thank Vincent D\'emery, Dominic Vella and Etienne Barthel for their help with the analytical and numerical analysis. This work was supported by the National Science Foundation under grant number DMR 1508186.
\end{acknowledgments}


\begin{thebibliography}{0}%
\makeatletter
\providecommand \@ifxundefined [1]{%
 \@ifx{#1\undefined}
}%
\providecommand \@ifnum [1]{%
 \ifnum #1\expandafter \@firstoftwo
 \else \expandafter \@secondoftwo
 \fi
}%
\providecommand \@ifx [1]{%
 \ifx #1\expandafter \@firstoftwo
 \else \expandafter \@secondoftwo
 \fi
}%
\providecommand \natexlab [1]{#1}%
\providecommand \enquote  [1]{``#1''}%
\providecommand \bibnamefont  [1]{#1}%
\providecommand \bibfnamefont [1]{#1}%
\providecommand \citenamefont [1]{#1}%
\providecommand \href@noop [0]{\@secondoftwo}%
\providecommand \href [0]{\begingroup \@sanitize@url \@href}%
\providecommand \@href[1]{\@@startlink{#1}\@@href}%
\providecommand \@@href[1]{\endgroup#1\@@endlink}%
\providecommand \@sanitize@url [0]{\catcode `\\12\catcode `\$12\catcode
  `\&12\catcode `\#12\catcode `\^12\catcode `\_12\catcode `\%12\relax}%
\providecommand \@@startlink[1]{}%
\providecommand \@@endlink[0]{}%
\providecommand \url  [0]{\begingroup\@sanitize@url \@url }%
\providecommand \@url [1]{\endgroup\@href {#1}{\urlprefix }}%
\providecommand \urlprefix  [0]{URL }%
\providecommand \Eprint [0]{\href }%
\providecommand \doibase [0]{http://dx.doi.org/}%
\providecommand \selectlanguage [0]{\@gobble}%
\providecommand \bibinfo  [0]{\@secondoftwo}%
\providecommand \bibfield  [0]{\@secondoftwo}%
\providecommand \translation [1]{[#1]}%
\providecommand \BibitemOpen [0]{}%
\providecommand \bibitemStop [0]{}%
\providecommand \bibitemNoStop [0]{.\EOS\space}%
\providecommand \EOS [0]{\spacefactor3000\relax}%
\providecommand \BibitemShut  [1]{\csname bibitem#1\endcsname}%
\let\auto@bib@innerbib\@empty
\end{thebibliography}%


\begin{thebibliography}{34}%
\makeatletter
\providecommand \@ifxundefined [1]{%
 \@ifx{#1\undefined}
}%
\providecommand \@ifnum [1]{%
 \ifnum #1\expandafter \@firstoftwo
 \else \expandafter \@secondoftwo
 \fi
}%
\providecommand \@ifx [1]{%
 \ifx #1\expandafter \@firstoftwo
 \else \expandafter \@secondoftwo
 \fi
}%
\providecommand \natexlab [1]{#1}%
\providecommand \enquote  [1]{``#1''}%
\providecommand \bibnamefont  [1]{#1}%
\providecommand \bibfnamefont [1]{#1}%
\providecommand \citenamefont [1]{#1}%
\providecommand \href@noop [0]{\@secondoftwo}%
\providecommand \href [0]{\begingroup \@sanitize@url \@href}%
\providecommand \@href[1]{\@@startlink{#1}\@@href}%
\providecommand \@@href[1]{\endgroup#1\@@endlink}%
\providecommand \@sanitize@url [0]{\catcode `\\12\catcode `\$12\catcode
  `\&12\catcode `\#12\catcode `\^12\catcode `\_12\catcode `\%12\relax}%
\providecommand \@@startlink[1]{}%
\providecommand \@@endlink[0]{}%
\providecommand \url  [0]{\begingroup\@sanitize@url \@url }%
\providecommand \@url [1]{\endgroup\@href {#1}{\urlprefix }}%
\providecommand \urlprefix  [0]{URL }%
\providecommand \Eprint [0]{\href }%
\providecommand \doibase [0]{http://dx.doi.org/}%
\providecommand \selectlanguage [0]{\@gobble}%
\providecommand \bibinfo  [0]{\@secondoftwo}%
\providecommand \bibfield  [0]{\@secondoftwo}%
\providecommand \translation [1]{[#1]}%
\providecommand \BibitemOpen [0]{}%
\providecommand \bibitemStop [0]{}%
\providecommand \bibitemNoStop [0]{.\EOS\space}%
\providecommand \EOS [0]{\spacefactor3000\relax}%
\providecommand \BibitemShut  [1]{\csname bibitem#1\endcsname}%
\let\auto@bib@innerbib\@empty
\bibitem [{\citenamefont {Gardel}\ \emph {et~al.}(2004)\citenamefont {Gardel},
  \citenamefont {Shin}, \citenamefont {MacKintosh}, \citenamefont {Mahadevan},
  \citenamefont {Matsudaira},\ and\ \citenamefont {Weitz}}]{Gardel2004}%
  \BibitemOpen
  \bibfield  {author} {\bibinfo {author} {\bibfnamefont {M.~L.}\ \bibnamefont
  {Gardel}}, \bibinfo {author} {\bibfnamefont {J.H.}\ \bibnamefont {Shin}},
  \bibinfo {author} {\bibfnamefont {F.C.}\ \bibnamefont {MacKintosh}}, \bibinfo
  {author} {\bibfnamefont {L.}~\bibnamefont {Mahadevan}}, \bibinfo {author}
  {\bibfnamefont {P.}~\bibnamefont {Matsudaira}}, \ and\ \bibinfo {author}
  {\bibfnamefont {D.A.}\ \bibnamefont {Weitz}},\ }\bibfield  {title} {\enquote
  {\bibinfo {title} {Elastic behavior of cross-linked and bundled actin
  networks},}\ }\href@noop {} {\bibfield  {journal} {\bibinfo  {journal}
  {Science}\ }\textbf {\bibinfo {volume} {304}},\ \bibinfo {pages} {1301}
  (\bibinfo {year} {2004})}\BibitemShut {NoStop}%
\bibitem [{\citenamefont {Chaudhuri}\ \emph {et~al.}(2007)\citenamefont
  {Chaudhuri}, \citenamefont {Parekh},\ and\ \citenamefont
  {Fletcher}}]{Chaudhuri2007}%
  \BibitemOpen
  \bibfield  {author} {\bibinfo {author} {\bibfnamefont {O.}~\bibnamefont
  {Chaudhuri}}, \bibinfo {author} {\bibfnamefont {S.~H.}\ \bibnamefont
  {Parekh}}, \ and\ \bibinfo {author} {\bibfnamefont {D.~A.}\ \bibnamefont
  {Fletcher}},\ }\bibfield  {title} {\enquote {\bibinfo {title} {Reversible
  stress softening of actin networks},}\ }\href@noop {} {\bibfield  {journal}
  {\bibinfo  {journal} {Nature}\ }\textbf {\bibinfo {volume} {445}},\ \bibinfo
  {pages} {295} (\bibinfo {year} {2007})}\BibitemShut {NoStop}%
\bibitem [{\citenamefont {Jiang}\ and\ \citenamefont
  {Zhang}(2008)}]{Jiang2008}%
  \BibitemOpen
  \bibfield  {author} {\bibinfo {author} {\bibfnamefont {H.}~\bibnamefont
  {Jiang}}\ and\ \bibinfo {author} {\bibfnamefont {J.}~\bibnamefont {Zhang}},\
  }\bibfield  {title} {\enquote {\bibinfo {title} {Mechanics of microtubule
  buckling supported by cytoplasm},}\ }\href@noop {} {\bibfield  {journal}
  {\bibinfo  {journal} {Journal of Applied Mechanics}\ }\textbf {\bibinfo
  {volume} {75}},\ \bibinfo {pages} {061019} (\bibinfo {year}
  {2008})}\BibitemShut {NoStop}%
\bibitem [{\citenamefont {Li}(2008)}]{Li2008}%
  \BibitemOpen
  \bibfield  {author} {\bibinfo {author} {\bibfnamefont {T.}~\bibnamefont
  {Li}},\ }\bibfield  {title} {\enquote {\bibinfo {title} {A mechanics model of
  microtubule buckling in living cells},}\ }\href@noop {} {\bibfield  {journal}
  {\bibinfo  {journal} {Journal of Biomechanics}\ }\textbf {\bibinfo {volume}
  {41}},\ \bibinfo {pages} {1722} (\bibinfo {year} {2008})}\BibitemShut
  {NoStop}%
\bibitem [{\citenamefont {Powers}(2010{\natexlab{a}})}]{Powers2010}%
  \BibitemOpen
  \bibfield  {author} {\bibinfo {author} {\bibfnamefont {Thomas~R.}\
  \bibnamefont {Powers}},\ }\bibfield  {title} {\enquote {\bibinfo {title}
  {Dynamics of filaments and membranes in a viscous fluid},}\ }\href@noop {}
  {\bibfield  {journal} {\bibinfo  {journal} {Review of Modern Physics}\
  }\textbf {\bibinfo {volume} {82}},\ \bibinfo {pages} {1607} (\bibinfo {year}
  {2010}{\natexlab{a}})}\BibitemShut {NoStop}%
\bibitem [{\citenamefont {Goldstein}\ and\ \citenamefont
  {Goriely}(2006)}]{Goldstein2006}%
  \BibitemOpen
  \bibfield  {author} {\bibinfo {author} {\bibfnamefont {Raymond~E.}\
  \bibnamefont {Goldstein}}\ and\ \bibinfo {author} {\bibfnamefont {Alain}\
  \bibnamefont {Goriely}},\ }\bibfield  {title} {\enquote {\bibinfo {title}
  {Dynamic buckling of morphoelastic filaments},}\ }\href@noop {} {\bibfield
  {journal} {\bibinfo  {journal} {Physical Review E: Rapid Communications}\
  }\textbf {\bibinfo {volume} {74}},\ \bibinfo {pages} {010901} (\bibinfo
  {year} {2006})}\BibitemShut {NoStop}%
\bibitem [{\citenamefont {Son}\ \emph {et~al.}(2013)\citenamefont {Son},
  \citenamefont {Guasto},\ and\ \citenamefont {Stocker}}]{Son2013}%
  \BibitemOpen
  \bibfield  {author} {\bibinfo {author} {\bibfnamefont {K.}~\bibnamefont
  {Son}}, \bibinfo {author} {\bibfnamefont {J.S.}\ \bibnamefont {Guasto}}, \
  and\ \bibinfo {author} {\bibfnamefont {R.}~\bibnamefont {Stocker}},\
  }\bibfield  {title} {\enquote {\bibinfo {title} {Bacteria can exploit a
  flagellar buckling instability to change direction},}\ }\href@noop {}
  {\bibfield  {journal} {\bibinfo  {journal} {Nature Physics}\ }\textbf
  {\bibinfo {volume} {9}},\ \bibinfo {pages} {494} (\bibinfo {year}
  {2013})}\BibitemShut {NoStop}%
\bibitem [{\citenamefont {Lindner}\ and\ \citenamefont
  {Shelley}(2012)}]{Lindner2012}%
  \BibitemOpen
  \bibfield  {author} {\bibinfo {author} {\bibfnamefont {Anke}\ \bibnamefont
  {Lindner}}\ and\ \bibinfo {author} {\bibfnamefont {Michael}\ \bibnamefont
  {Shelley}},\ }\bibfield  {title} {\enquote {\bibinfo {title} {Elastic fibers
  in flows},}\ }\href@noop {} {\bibfield  {journal} {\bibinfo  {journal} {RSC
  Soft Matter No. 1}\ }\textbf {\bibinfo {volume} {82}},\ \bibinfo {pages}
  {1607} (\bibinfo {year} {2012})}\BibitemShut {NoStop}%
\bibitem [{\citenamefont {Biot}(1961)}]{Biot1961}%
  \BibitemOpen
  \bibfield  {author} {\bibinfo {author} {\bibfnamefont {M.~A.}\ \bibnamefont
  {Biot}},\ }\bibfield  {title} {\enquote {\bibinfo {title} {Theory of folding
  of stratified viscoelastic media and its implications in tectonics and
  orogenesis},}\ }\href@noop {} {\bibfield  {journal} {\bibinfo  {journal}
  {Geophysical Society of America Bulletin}\ }\textbf {\bibinfo {volume}
  {72}},\ \bibinfo {pages} {1595} (\bibinfo {year} {1961})}\BibitemShut
  {NoStop}%
\bibitem [{\citenamefont {Chopin}\ and\ \citenamefont
  {Kudrolli}(2013)}]{Chopin2013}%
  \BibitemOpen
  \bibfield  {author} {\bibinfo {author} {\bibfnamefont {J.}~\bibnamefont
  {Chopin}}\ and\ \bibinfo {author} {\bibfnamefont {A.}~\bibnamefont
  {Kudrolli}},\ }\bibfield  {title} {\enquote {\bibinfo {title} {Helicoids,
  wrinkles, and loops in twisted ribbons},}\ }\href@noop {} {\bibfield
  {journal} {\bibinfo  {journal} {Physical Review Letters}\ }\textbf {\bibinfo
  {volume} {111}},\ \bibinfo {pages} {174302} (\bibinfo {year}
  {2013})}\BibitemShut {NoStop}%
\bibitem [{\citenamefont {Chopin}\ \emph {et~al.}(2015)\citenamefont {Chopin},
  \citenamefont {Demery},\ and\ \citenamefont {Davidovitch}}]{Chopin2015}%
  \BibitemOpen
  \bibfield  {author} {\bibinfo {author} {\bibfnamefont {J.}~\bibnamefont
  {Chopin}}, \bibinfo {author} {\bibfnamefont {V.}~\bibnamefont {Demery}}, \
  and\ \bibinfo {author} {\bibfnamefont {B.}~\bibnamefont {Davidovitch}},\
  }\bibfield  {title} {\enquote {\bibinfo {title} {Roadmap to the morphological
  instabilities of a stretched twisted ribbon},}\ }\href@noop {} {\bibfield
  {journal} {\bibinfo  {journal} {Journal of Elasticity}\ }\textbf {\bibinfo
  {volume} {119}},\ \bibinfo {pages} {137--189} (\bibinfo {year}
  {2015})}\BibitemShut {NoStop}%
\bibitem [{\citenamefont {Miller}\ \emph {et~al.}(2015)\citenamefont {Miller},
  \citenamefont {Su}, \citenamefont {Pabon}, \citenamefont {Wicks},
  \citenamefont {Bertoldi},\ and\ \citenamefont {Reis}}]{miller2015buckling}%
  \BibitemOpen
  \bibfield  {author} {\bibinfo {author} {\bibfnamefont {JT}~\bibnamefont
  {Miller}}, \bibinfo {author} {\bibfnamefont {T}~\bibnamefont {Su}}, \bibinfo
  {author} {\bibfnamefont {J}~\bibnamefont {Pabon}}, \bibinfo {author}
  {\bibfnamefont {N}~\bibnamefont {Wicks}}, \bibinfo {author} {\bibfnamefont
  {K}~\bibnamefont {Bertoldi}}, \ and\ \bibinfo {author} {\bibfnamefont
  {PM}~\bibnamefont {Reis}},\ }\bibfield  {title} {\enquote {\bibinfo {title}
  {Buckling of a thin elastic rod inside a horizontal cylindrical
  constraint},}\ }\href@noop {} {\bibfield  {journal} {\bibinfo  {journal}
  {Extreme Mechanics Letters}\ }\textbf {\bibinfo {volume} {3}},\ \bibinfo
  {pages} {36--44} (\bibinfo {year} {2015})}\BibitemShut {NoStop}%
\bibitem [{\citenamefont {Vermorel}\ \emph {et~al.}(2007)\citenamefont
  {Vermorel}, \citenamefont {Vanderberghe},\ and\ \citenamefont
  {Villermaux}}]{Vermorel2007}%
  \BibitemOpen
  \bibfield  {author} {\bibinfo {author} {\bibfnamefont {R.}~\bibnamefont
  {Vermorel}}, \bibinfo {author} {\bibfnamefont {N.}~\bibnamefont
  {Vanderberghe}}, \ and\ \bibinfo {author} {\bibfnamefont {E.}~\bibnamefont
  {Villermaux}},\ }\bibfield  {title} {\enquote {\bibinfo {title} {Rubber band
  recoil},}\ }\href@noop {} {\bibfield  {journal} {\bibinfo  {journal}
  {Proceedings of the royal society London A}\ }\textbf {\bibinfo {volume}
  {463}},\ \bibinfo {pages} {641} (\bibinfo {year} {2007})}\BibitemShut
  {NoStop}%
\bibitem [{\citenamefont {Huang}\ and\ \citenamefont {Suo}(2001)}]{Huang2001}%
  \BibitemOpen
  \bibfield  {author} {\bibinfo {author} {\bibfnamefont {R.}~\bibnamefont
  {Huang}}\ and\ \bibinfo {author} {\bibfnamefont {Z.}~\bibnamefont {Suo}},\
  }\bibfield  {title} {\enquote {\bibinfo {title} {Wrinkling of a compressed
  elastic film on a viscous layer},}\ }\href@noop {} {\bibfield  {journal}
  {\bibinfo  {journal} {Journal of Applied Physics}\ }\textbf {\bibinfo
  {volume} {91}},\ \bibinfo {pages} {1135} (\bibinfo {year}
  {2001})}\BibitemShut {NoStop}%
\bibitem [{\citenamefont {Kodio}\ \emph {et~al.}(2016)\citenamefont {Kodio},
  \citenamefont {Griffiths},\ and\ \citenamefont
  {Vella}}]{kodio2016lubricated}%
  \BibitemOpen
  \bibfield  {author} {\bibinfo {author} {\bibfnamefont {Ousmane}\ \bibnamefont
  {Kodio}}, \bibinfo {author} {\bibfnamefont {Ian~M}\ \bibnamefont
  {Griffiths}}, \ and\ \bibinfo {author} {\bibfnamefont {Dominic}\ \bibnamefont
  {Vella}},\ }\bibfield  {title} {\enquote {\bibinfo {title} {Lubricated
  wrinkles: imposed constraints affect the dynamics of wrinkle coarsening},}\
  }\href@noop {} {\bibfield  {journal} {\bibinfo  {journal} {arXiv preprint
  arXiv:1609.04598}\ } (\bibinfo {year} {2016})}\BibitemShut {NoStop}%
\bibitem [{\citenamefont {Forterre}\ \emph {et~al.}(2005)\citenamefont
  {Forterre}, \citenamefont {Skotheim}, \citenamefont {Dumais},\ and\
  \citenamefont {Mahadevan}}]{forterre2005venus}%
  \BibitemOpen
  \bibfield  {author} {\bibinfo {author} {\bibfnamefont {Yo{\"e}l}\
  \bibnamefont {Forterre}}, \bibinfo {author} {\bibfnamefont {Jan~M}\
  \bibnamefont {Skotheim}}, \bibinfo {author} {\bibfnamefont {Jacques}\
  \bibnamefont {Dumais}}, \ and\ \bibinfo {author} {\bibfnamefont
  {Lakshminarayanan}\ \bibnamefont {Mahadevan}},\ }\bibfield  {title} {\enquote
  {\bibinfo {title} {How the venus flytrap snaps},}\ }\href@noop {} {\bibfield
  {journal} {\bibinfo  {journal} {Nature}\ }\textbf {\bibinfo {volume} {433}},\
  \bibinfo {pages} {421--425} (\bibinfo {year} {2005})}\BibitemShut {NoStop}%
\bibitem [{\citenamefont {Stuart}\ \emph {et~al.}(2010)\citenamefont {Stuart},
  \citenamefont {Huck}, \citenamefont {Genzer}, \citenamefont {M{\"u}ller},
  \citenamefont {Ober}, \citenamefont {Stamm}, \citenamefont {Sukhorukov},
  \citenamefont {Szleifer}, \citenamefont {Tsukruk}, \citenamefont {Urban}
  \emph {et~al.}}]{stuart2010emerging}%
  \BibitemOpen
  \bibfield  {author} {\bibinfo {author} {\bibfnamefont {Martien A~Cohen}\
  \bibnamefont {Stuart}}, \bibinfo {author} {\bibfnamefont {Wilhelm~TS}\
  \bibnamefont {Huck}}, \bibinfo {author} {\bibfnamefont {Jan}\ \bibnamefont
  {Genzer}}, \bibinfo {author} {\bibfnamefont {Marcus}\ \bibnamefont
  {M{\"u}ller}}, \bibinfo {author} {\bibfnamefont {Christopher}\ \bibnamefont
  {Ober}}, \bibinfo {author} {\bibfnamefont {Manfred}\ \bibnamefont {Stamm}},
  \bibinfo {author} {\bibfnamefont {Gleb~B}\ \bibnamefont {Sukhorukov}},
  \bibinfo {author} {\bibfnamefont {Igal}\ \bibnamefont {Szleifer}}, \bibinfo
  {author} {\bibfnamefont {Vladimir~V}\ \bibnamefont {Tsukruk}}, \bibinfo
  {author} {\bibfnamefont {Marek}\ \bibnamefont {Urban}},  \emph {et~al.},\
  }\bibfield  {title} {\enquote {\bibinfo {title} {Emerging applications of
  stimuli-responsive polymer materials},}\ }\href@noop {} {\bibfield  {journal}
  {\bibinfo  {journal} {Nature materials}\ }\textbf {\bibinfo {volume} {9}},\
  \bibinfo {pages} {101--113} (\bibinfo {year} {2010})}\BibitemShut {NoStop}%
\bibitem [{\citenamefont {Jager}\ \emph {et~al.}(2000)\citenamefont {Jager},
  \citenamefont {Smela},\ and\ \citenamefont
  {Ingan{\"a}s}}]{jager2000microfabricating}%
  \BibitemOpen
  \bibfield  {author} {\bibinfo {author} {\bibfnamefont {Edwin~WH}\
  \bibnamefont {Jager}}, \bibinfo {author} {\bibfnamefont {Elisabeth}\
  \bibnamefont {Smela}}, \ and\ \bibinfo {author} {\bibfnamefont {Olle}\
  \bibnamefont {Ingan{\"a}s}},\ }\bibfield  {title} {\enquote {\bibinfo {title}
  {Microfabricating conjugated polymer actuators},}\ }\href@noop {} {\bibfield
  {journal} {\bibinfo  {journal} {Science}\ }\textbf {\bibinfo {volume}
  {290}},\ \bibinfo {pages} {1540--1545} (\bibinfo {year} {2000})}\BibitemShut
  {NoStop}%
\bibitem [{\citenamefont {Osada}\ \emph {et~al.}(1992)\citenamefont {Osada},
  \citenamefont {Okuzaki},\ and\ \citenamefont {Hori}}]{osada1992polymer}%
  \BibitemOpen
  \bibfield  {author} {\bibinfo {author} {\bibfnamefont {Yoshihito}\
  \bibnamefont {Osada}}, \bibinfo {author} {\bibfnamefont {Hidenori}\
  \bibnamefont {Okuzaki}}, \ and\ \bibinfo {author} {\bibfnamefont {Hirofumi}\
  \bibnamefont {Hori}},\ }\bibfield  {title} {\enquote {\bibinfo {title} {A
  polymer gel with electrically driven motility},}\ }\href@noop {} {\bibfield
  {journal} {\bibinfo  {journal} {Nature}\ }\textbf {\bibinfo {volume} {355}},\
  \bibinfo {pages} {242--244} (\bibinfo {year} {1992})}\BibitemShut {NoStop}%
\bibitem [{\citenamefont {Chen}\ and\ \citenamefont
  {Hoffman}(1995)}]{chen1995graft}%
  \BibitemOpen
  \bibfield  {author} {\bibinfo {author} {\bibfnamefont {Guohua}\ \bibnamefont
  {Chen}}\ and\ \bibinfo {author} {\bibfnamefont {Allan~S}\ \bibnamefont
  {Hoffman}},\ }\bibfield  {title} {\enquote {\bibinfo {title} {Graft
  copolymers that exhibit temperature-induced phase transitions over a wide
  range of ph},}\ }\href@noop {} {\bibfield  {journal} {\bibinfo  {journal}
  {Nature}\ }\textbf {\bibinfo {volume} {373}},\ \bibinfo {pages} {49--52}
  (\bibinfo {year} {1995})}\BibitemShut {NoStop}%
\bibitem [{\citenamefont {Kim}\ \emph {et~al.}(2012)\citenamefont {Kim},
  \citenamefont {Hanna}, \citenamefont {Byun}, \citenamefont {Santangelo},\
  and\ \citenamefont {Hayward}}]{kim2012designing}%
  \BibitemOpen
  \bibfield  {author} {\bibinfo {author} {\bibfnamefont {Jungwook}\
  \bibnamefont {Kim}}, \bibinfo {author} {\bibfnamefont {James~A}\ \bibnamefont
  {Hanna}}, \bibinfo {author} {\bibfnamefont {Myunghwan}\ \bibnamefont {Byun}},
  \bibinfo {author} {\bibfnamefont {Christian~D}\ \bibnamefont {Santangelo}}, \
  and\ \bibinfo {author} {\bibfnamefont {Ryan~C}\ \bibnamefont {Hayward}},\
  }\bibfield  {title} {\enquote {\bibinfo {title} {Designing responsive buckled
  surfaces by halftone gel lithography},}\ }\href@noop {} {\bibfield  {journal}
  {\bibinfo  {journal} {Science}\ }\textbf {\bibinfo {volume} {335}},\ \bibinfo
  {pages} {1201--1205} (\bibinfo {year} {2012})}\BibitemShut {NoStop}%
\bibitem [{\citenamefont {Camacho-Lopez}\ \emph {et~al.}(2004)\citenamefont
  {Camacho-Lopez}, \citenamefont {Finkelmann}, \citenamefont {Palffy-Muhoray},\
  and\ \citenamefont {Shelley}}]{camacho2004fast}%
  \BibitemOpen
  \bibfield  {author} {\bibinfo {author} {\bibfnamefont {Miguel}\ \bibnamefont
  {Camacho-Lopez}}, \bibinfo {author} {\bibfnamefont {Heino}\ \bibnamefont
  {Finkelmann}}, \bibinfo {author} {\bibfnamefont {Peter}\ \bibnamefont
  {Palffy-Muhoray}}, \ and\ \bibinfo {author} {\bibfnamefont {Michael}\
  \bibnamefont {Shelley}},\ }\bibfield  {title} {\enquote {\bibinfo {title}
  {Fast liquid-crystal elastomer swims into the dark},}\ }\href@noop {}
  {\bibfield  {journal} {\bibinfo  {journal} {Nature Materials}\ }\textbf
  {\bibinfo {volume} {3}},\ \bibinfo {pages} {307--310} (\bibinfo {year}
  {2004})}\BibitemShut {NoStop}%
\bibitem [{\citenamefont {van Oosten}\ \emph {et~al.}(2009)\citenamefont {van
  Oosten}, \citenamefont {Bastiaansen},\ and\ \citenamefont
  {Broer}}]{van2009printed}%
  \BibitemOpen
  \bibfield  {author} {\bibinfo {author} {\bibfnamefont {Casper~L}\
  \bibnamefont {van Oosten}}, \bibinfo {author} {\bibfnamefont {Cees~WM}\
  \bibnamefont {Bastiaansen}}, \ and\ \bibinfo {author} {\bibfnamefont
  {Dirk~J}\ \bibnamefont {Broer}},\ }\bibfield  {title} {\enquote {\bibinfo
  {title} {Printed artificial cilia from liquid-crystal network actuators
  modularly driven by light},}\ }\href@noop {} {\bibfield  {journal} {\bibinfo
  {journal} {Nature Materials}\ }\textbf {\bibinfo {volume} {8}},\ \bibinfo
  {pages} {677--682} (\bibinfo {year} {2009})}\BibitemShut {NoStop}%
\bibitem [{\citenamefont {Yu}\ \emph {et~al.}(2003)\citenamefont {Yu},
  \citenamefont {Nakano},\ and\ \citenamefont {Ikeda}}]{yu2003photomechanics}%
  \BibitemOpen
  \bibfield  {author} {\bibinfo {author} {\bibfnamefont {Yanlei}\ \bibnamefont
  {Yu}}, \bibinfo {author} {\bibfnamefont {Makoto}\ \bibnamefont {Nakano}}, \
  and\ \bibinfo {author} {\bibfnamefont {Tomiki}\ \bibnamefont {Ikeda}},\
  }\bibfield  {title} {\enquote {\bibinfo {title} {Photomechanics: directed
  bending of a polymer film by light},}\ }\href@noop {} {\bibfield  {journal}
  {\bibinfo  {journal} {Nature}\ }\textbf {\bibinfo {volume} {425}},\ \bibinfo
  {pages} {145--145} (\bibinfo {year} {2003})}\BibitemShut {NoStop}%
\bibitem [{\citenamefont {Gladden}\ \emph {et~al.}(2005)\citenamefont
  {Gladden}, \citenamefont {Handzy}, \citenamefont {Belmonte},\ and\
  \citenamefont {Villermaux}}]{gladden2005dynamic}%
  \BibitemOpen
  \bibfield  {author} {\bibinfo {author} {\bibfnamefont {JR}~\bibnamefont
  {Gladden}}, \bibinfo {author} {\bibfnamefont {NZ}~\bibnamefont {Handzy}},
  \bibinfo {author} {\bibfnamefont {Andrew}\ \bibnamefont {Belmonte}}, \ and\
  \bibinfo {author} {\bibfnamefont {Emmanuel}\ \bibnamefont {Villermaux}},\
  }\bibfield  {title} {\enquote {\bibinfo {title} {Dynamic buckling and
  fragmentation in brittle rods},}\ }\href@noop {} {\bibfield  {journal}
  {\bibinfo  {journal} {Physical Review Letters}\ }\textbf {\bibinfo {volume}
  {94}},\ \bibinfo {pages} {035503} (\bibinfo {year} {2005})}\BibitemShut
  {NoStop}%
\bibitem [{\citenamefont {Brangwynne}\ \emph {et~al.}(2007)\citenamefont
  {Brangwynne}, \citenamefont {Koenderink}, \citenamefont {Barry},
  \citenamefont {Dogic}, \citenamefont {MacKintosh},\ and\ \citenamefont
  {Weitz}}]{Brangwynne2007}%
  \BibitemOpen
  \bibfield  {author} {\bibinfo {author} {\bibfnamefont {C.P.}\ \bibnamefont
  {Brangwynne}}, \bibinfo {author} {\bibfnamefont {G.H.}\ \bibnamefont
  {Koenderink}}, \bibinfo {author} {\bibfnamefont {E.}~\bibnamefont {Barry}},
  \bibinfo {author} {\bibfnamefont {Z.}~\bibnamefont {Dogic}}, \bibinfo
  {author} {\bibfnamefont {F.C.}\ \bibnamefont {MacKintosh}}, \ and\ \bibinfo
  {author} {\bibfnamefont {David~A.}\ \bibnamefont {Weitz}},\ }\bibfield
  {title} {\enquote {\bibinfo {title} {Bending dynamics of fluctuating
  biopolymers probed by automated high-resolution filament tracking},}\
  }\href@noop {} {\bibfield  {journal} {\bibinfo  {journal} {Biophysical
  Journal}\ }\textbf {\bibinfo {volume} {93}},\ \bibinfo {pages} {346}
  (\bibinfo {year} {2007})}\BibitemShut {NoStop}%
\bibitem [{\citenamefont {Wiggins}\ \emph {et~al.}(1998)\citenamefont
  {Wiggins}, \citenamefont {Riveline}, \citenamefont {Ott},\ and\ \citenamefont
  {Goldstein}}]{wiggins1998trapping}%
  \BibitemOpen
  \bibfield  {author} {\bibinfo {author} {\bibfnamefont {Chris~H}\ \bibnamefont
  {Wiggins}}, \bibinfo {author} {\bibfnamefont {D}~\bibnamefont {Riveline}},
  \bibinfo {author} {\bibfnamefont {Albrecht}\ \bibnamefont {Ott}}, \ and\
  \bibinfo {author} {\bibfnamefont {Raymond~E}\ \bibnamefont {Goldstein}},\
  }\bibfield  {title} {\enquote {\bibinfo {title} {Trapping and wiggling:
  elastohydrodynamics of driven microfilaments},}\ }\href@noop {} {\bibfield
  {journal} {\bibinfo  {journal} {Biophysical Journal}\ }\textbf {\bibinfo
  {volume} {74}},\ \bibinfo {pages} {1043--1060} (\bibinfo {year}
  {1998})}\BibitemShut {NoStop}%
\bibitem [{\citenamefont {Batchelor}(1970)}]{batchelor1970slender}%
  \BibitemOpen
  \bibfield  {author} {\bibinfo {author} {\bibfnamefont {GK}~\bibnamefont
  {Batchelor}},\ }\bibfield  {title} {\enquote {\bibinfo {title} {Slender-body
  theory for particles of arbitrary cross-section in stokes flow},}\
  }\href@noop {} {\bibfield  {journal} {\bibinfo  {journal} {Journal of Fluid
  Mechanics}\ }\textbf {\bibinfo {volume} {44}},\ \bibinfo {pages} {419--440}
  (\bibinfo {year} {1970})}\BibitemShut {NoStop}%
\bibitem [{\citenamefont {Powers}(2010{\natexlab{b}})}]{powers2010dynamics}%
  \BibitemOpen
  \bibfield  {author} {\bibinfo {author} {\bibfnamefont {Thomas~R}\
  \bibnamefont {Powers}},\ }\bibfield  {title} {\enquote {\bibinfo {title}
  {Dynamics of filaments and membranes in a viscous fluid},}\ }\href@noop {}
  {\bibfield  {journal} {\bibinfo  {journal} {Reviews of Modern Physics}\
  }\textbf {\bibinfo {volume} {82}},\ \bibinfo {pages} {1607} (\bibinfo {year}
  {2010}{\natexlab{b}})}\BibitemShut {NoStop}%
\bibitem [{\citenamefont {Landau}\ and\ \citenamefont
  {Lifshitz}(1986)}]{landau1986course}%
  \BibitemOpen
  \bibfield  {author} {\bibinfo {author} {\bibfnamefont {Lev~Davidovich}\
  \bibnamefont {Landau}}\ and\ \bibinfo {author} {\bibfnamefont {Eugin~M}\
  \bibnamefont {Lifshitz}},\ }\href@noop {} {\emph {\bibinfo {title} {Course of
  theoretical physics, Theory of elasticity}}}\ (\bibinfo  {publisher}
  {Pergamon Press Oxford},\ \bibinfo {year} {1986})\BibitemShut {NoStop}%
\bibitem [{Note1()}]{Note1}%
  \BibitemOpen
  \bibinfo {note} {See Supplementary Documentation for a movie of the shape
  evolution, the calculation of the Euler buckling modes, and the comparison of
  the fundamental mode with the final shape shown of the filament.}\BibitemShut
  {Stop}%
\bibitem [{\citenamefont {Gosselin}\ \emph {et~al.}(2014)\citenamefont
  {Gosselin}, \citenamefont {Neetzow},\ and\ \citenamefont
  {Paak}}]{Gosselin2014}%
  \BibitemOpen
  \bibfield  {author} {\bibinfo {author} {\bibfnamefont {F.~P.}\ \bibnamefont
  {Gosselin}}, \bibinfo {author} {\bibfnamefont {P.}~\bibnamefont {Neetzow}}, \
  and\ \bibinfo {author} {\bibfnamefont {M.}~\bibnamefont {Paak}},\ }\bibfield
  {title} {\enquote {\bibinfo {title} {Buckling of a beam extruded into highly
  viscous fluid},}\ }\href@noop {} {\bibfield  {journal} {\bibinfo  {journal}
  {Physical Review E}\ }\textbf {\bibinfo {volume} {90}},\ \bibinfo {pages}
  {052718} (\bibinfo {year} {2014})}\BibitemShut {NoStop}%
\bibitem [{\citenamefont {Wouwer}\ \emph {et~al.}(2014)\citenamefont {Wouwer},
  \citenamefont {Saucez}, \citenamefont {Vilas} \emph
  {et~al.}}]{wouwer2014simulation}%
  \BibitemOpen
  \bibfield  {author} {\bibinfo {author} {\bibfnamefont {Alain~Vande}\
  \bibnamefont {Wouwer}}, \bibinfo {author} {\bibfnamefont {Philippe}\
  \bibnamefont {Saucez}}, \bibinfo {author} {\bibfnamefont {Carlos}\
  \bibnamefont {Vilas}},  \emph {et~al.},\ }\href@noop {} {\emph {\bibinfo
  {title} {Simulation of ODE/PDE Models with MATLAB, OCTAVE and SCILAB}}}\
  (\bibinfo  {publisher} {Springer},\ \bibinfo {year} {2014})\BibitemShut
  {NoStop}%
\bibitem [{\citenamefont {Budd}\ and\ \citenamefont
  {Peletier}(2000)}]{budd2000approximate}%
  \BibitemOpen
  \bibfield  {author} {\bibinfo {author} {\bibfnamefont {Chris~J}\ \bibnamefont
  {Budd}}\ and\ \bibinfo {author} {\bibfnamefont {Mark~A}\ \bibnamefont
  {Peletier}},\ }\bibfield  {title} {\enquote {\bibinfo {title} {Approximate
  self-similarity in models of geological folding},}\ }\href@noop {} {\bibfield
   {journal} {\bibinfo  {journal} {SIAM Journal on Applied Mathematics}\
  }\textbf {\bibinfo {volume} {60}},\ \bibinfo {pages} {990--1016} (\bibinfo
  {year} {2000})}\BibitemShut {NoStop}%
\end{thebibliography}%
\end{document}